\documentstyle[12pt]{article}

\newcommand{\beq}{\begin{equation}}
\newcommand{\eeq}{\end{equation}}
\input{tcilatex}
\begin{document}

\begin{titlepage}
\begin{center}
{\hbox to\hsize{ 
\hfill \bf HIP-2001-04/TH}}
{\hbox to\hsize{
\hfill \bf {\bf hep-th/0102207}}}


\vspace{2.5cm}

{\Large \bf  
Vanishing of cosmological constant and fully localized gravity 
in a Brane World with extra time(s) 
\\}

\vspace{1.5cm}

{\bf  Z. Berezhiani$^{\mathrm{a,c}}$, M. Chaichian$^{\mathrm{b}}$, A.B.
Kobakhidze$^{\mathrm{b,c}}$ and Z.-H. Yu$^{\mathrm{b}}$   \\}

\smallskip

{ \small \it  
$^{\mathrm{a}}$Dipartamento di Fisica, Universit\'{a} di L'Aquila, I-67010
Coppito, AQ, and \\
INFN, Laboratori Nazionali del Gran Sasso, I-67010 Assergi, AQ, Italy \\
$^{\mathrm{b}}$HEP Division, Department of Physics, University of Helsinki
and \\ Helsinki Institute of Physics, FIN-00014 Helsinki, Finland\\
$^{\mathrm{c}}$Andronikashvili Institute of Physics,GE-380077 Tbilisi, Georgia. }

\vspace{1.5cm}

{\bf Abstract}\\
\end{center}
\noindent
We construct an intersecting brane configuration in six-dimensional space 
with one extra space-like and one extra time-like dimensions. With a certain
additional symmetry imposed on the extra space-time we have found that effective
four-dimensional cosmological constant vanishes automatically, providing the static 
solution with gravity fully localized at the intersection region as   
there are no propagating  
massive modes of graviton. In this way, the same symmetry
allows us to eliminate tachyonic states of graviton from the spectrum of the effective  
four-dimensional theory, thus avoiding phenomenological difficulties
comming from the matter instability usually induced in theories with extra time-like 
dimensions.

\end{titlepage}
\baselineskip=16pt

\section{Introduction.}

Recently the idea that our world is confined on the four-dimensional
subspace of higher dimensional space-time \cite{1} attracts considerable
attention. Within this scenario, often known as a ''Brane World'' scenario,
several phenomenological, cosmological and astrophysical issues have been
revisited. What is particularly exiting within this approach is that the
size of extra dimensions can be large \cite{2} or even infinite \cite{3,4,5}%
, leading to the interesting theoretical alternative to conventional
Kaluza-Klein (KK) compactification as well as to the specific predictions
that can be tested experimentally in the visible future.

One of the original motivations for the recent versions of Brane World
models was an explanation of the apparent hierarchy among two fundamental
scales, the Planck scale $M_{Pl}$, and the electroweak scale $M_{W}$ \cite
{2,6} (see also \cite{7}). The scenario of Refs. \cite{2} utilize $\delta $
extra compact dimensions with large compactification radii $r_{n}$ ($%
n=1,...,\delta $) in the factorizable, $M^{4}\times N^{\delta }$, ($4+\delta 
$)-dimensional space-time and thus the apparent weakness of gravity in the
visible four-dimensional world ($M^{4}$, 3-brane) is explained due to the
large volume $V_{N^{\delta }}\sim \prod_{n=1}^{\delta }r_{n}$ of the
extra-dimensional submanifold $N^{\delta }$ : 
\begin{equation}
M_{Pl}^{2}=M_{*}^{\delta +2}V_{N^{\delta }},  \label{1}
\end{equation}
where $M_{*}$ is the fundamental high-dimensional scale and $M_{Pl}$ is the
ordinary four-dimensional Planck scale.

The scenario of Ref. \cite{6} deals with a 5-dimensional non-factorizable AdS%
$_{5}$ space-time with two 3-branes located at the $S^{1}/Z_{2}$ orbifold
fixed points of the fifth compact dimension. Now the weakness of gravity in
the visible world 3-brane is explained without recourse to large extra
dimensions, but rather as a result of gravity localization on the hidden
3-brane. Gravity localization in such scenario occurs because the
five-dimensional Einstein's equations admit the solution for the space-time
metric with a scale factor (``warp factor'') which is a falling exponential
function of the distance along the extra dimension y perpendicular to the
branes\footnote{%
Generalizations to higher-dimensional space-times is possible within the
intersecting branes scenarios \cite{8}, models with string-like defects (in
the case of 2 extra dimensions) \cite{9} or along the lines discussed in 
\cite{10}.}: 
\begin{equation}
ds^{2}=e^{-2k|y|}dx_{1+3}^{2}+dy^{2},  \label{2}
\end{equation}
when the bulk cosmological constant $\Lambda $ ($\Lambda <0$) and the
tensions $T_{vis}$ and $T_{hid}$ of the visible and hidden branes
respectively are related according to: 
\begin{equation}
T_{hid}=-T_{vis}=6M_{*}^{3}k,~~~k=\sqrt{-\frac{\Lambda }{6M_{*}^{3}}}.
\label{3}
\end{equation}
Thus, graviton is essentially localized on the hidden brane with positive
tension ($T_{hid}>0$) which is located at $y=0$ fixed point of the $%
S^{1}/Z_{2}$ orbifold, while the Standard Model particles are assumed to be
restricted on the visible brane with negative tension ($T_{vis}<0$) which is
located at $y=\pi r_{c}$ ($r_{c}$ is the size of extra dimension) orbifold
fixed point. So, a hierarchically small scale factor generated for the
metric on the visible brane gives an exponential hierarchy between the mass
scales of the visible brane and the fundamental mass scale $M_{*}$, after
one appropriately rescales the fields on the visible brane. In fact,
assuming $M_{*}\sim M_{Pl}$, TeV-sized electroweak scale can be generated on
the visible brane by requiring $r_{c}\cdot M_{*}\simeq 12$.

However, there is even more severe hierarchy problem afflicting fundamental
physics which is related to the observed smallness of the 4-dimensional
cosmological constant (for reviews see \cite{11}). In the original proposals
with large \cite{2} or warped \cite{3,6} extra dimensions this problem
remains untouched. Say, the fine-tuned condition (\ref{3}) which provides
the solution (\ref{2}) is nothing but the condition for the vanishing of the
effective 4-dimensional cosmological constant. Several more or less
successful attempts have been made recently to attack this problem within
the Brane World models \cite{12,13,14,15,16,17,18}. One solution is to
ensure dynamical self-adjustment of the relation (\ref{3}) by introducing an
extra bulk scalar field $\phi $ with an appropriate bulk potential \cite{12}%
. In this scenario, however, the scalar fields $\phi $ becomes singular at a
finite distance along the extra dimension\footnote{%
For a non-singular solution see \cite{13}.} and the warp factor in the
metric (\ref{2}) vanishes at singularity. As a result the whole background
solution becomes unstable under the bulk perturbations and any procedure
which regularize the singularity reintroduces the fine-tuning back.\cite{14}%
. It was shown also in \cite{15} that non-static (cosmological) solutions of 
\cite{12} might be unstable as well, thus leading to the energy
non-conservation as brane world expands (contracts).

In this paper we would like to suggest an alternative possibility to
overcome an unnatural fine-tuning (\ref{3}) by introducing an extra
time-like dimensions. Particularly, we construct the configuration of
intersecting branes in a 6-dimensional bulk space-time with signature $(4,2)$%
. With a certain additional symmetry imposed on the extra space-time, we
find that effective 4-dimensional cosmological constant automatically
vanishes, providing a static solution with gravity localized at the
intersection region. Remarkable, the symmetry which forced the cosmological
constant to vanish prevents, at the same time, propagation of a massive
Kaluza-Klein (KK) modes of graviton (including those of tachyonic) and thus
gives the full localization of gravity at the intersection of branes, while
non-trivial warp factor of the background metric allows us to solve
Planck/TeV scale hierarchy problem by placing visible 3-brane at distances $%
{\cal O}(10^{2}M_{Pl})$ away from the intersection. 

\section{Extra time-like dimensions and null space dimensional reduction.}

Except of a very few recent works \cite{19,20,21,22,23,24,k1} (see also \cite
{25,26,27} for some earlier works on extra time-like dimensions), most of
the Brane World models considered so far deal with extra space-like
dimensions. The reason is some pathological features of field theories with
extra time-like dimensions associated with appearance of ghost and tachyonic
states in the effective four dimensional theory and related with them
instabilities of various kind \cite{19,20,23,24,26}. However, there are no
firm theoretical reasons why extra time-like dimensions cannot exist and,
moreover, they indeed appear at the fundamental level within the various
versions of string theory (see e.g. \cite{27} for some recent approaches).
The tachyonic instabilities, being the subject of experimental verification,
can be accommodated within the theoretical models if they do not contradict
the existing experimental date. Indeed, in the case of conventional KK
compactification with factorized geometry of space-time one can satisfy
experimental upper limits on various processes leading to the instability of
matter by choosing sufficiently small radii for the compact extra time-like
dimensions \cite{19, 22}\footnote{%
It is interesting that in the case when only gravity feels extra times \cite
{19,22} the imaginary part of the gravitational self-energy due to the
exchange of tachyonic KK gravitons of the spherically symmetric body is
periodic and for some critical radii of the body $R=2\pi kL$ (where $L$ is a
size of extra time-like dimension and $k\in N$) it vanishes, so such
self-gravitating body in fact is stable. Another interesting phenomena
observed in \cite{19,22} is the screening of the gravitational force due to
the contribution of the tachyonic KK gravitons.}. The same is actually true
for the case of warped compactification \cite{3,6} if our visible world is
confined on the brane which is sufficiently close to the so called Planck
scale brane where the graviton zero mode is localized. Otherwise, demanding
that the visible brane is displaced from the Placnk brane at distances which
provides the solution to the Planck/TeV hierarchy problem, one gets for
instance an unacceptable rapid gravitational decay of neutron \cite{20,24}.
Thus, it seems that generation of Planck/TeV hierarchy is incompatible with
extra time-like dimensions in both scenarios with factorizable \cite{19} and
non-factorizable geometry \cite{20}.

Of course, it is more desirable to overcome appearance of ghost and tachyons
in the four-dimensional effective theory. One obvious way to do this is to
choose a certain geometry of the extra space preventing the appearance of
ghost and tachyons. Indeed, for example, one can start with a topological
gravity in higher dimensions generating usual Einstein's gravity in
four-dimensional subspace \cite{21}. Then, since the bulk space-time is
topological the propagation of gravitons in the extra space does not take
place at all and thus tachyonic KK gravitons do not appear in the effective
four-dimensional theory. In such a case, however extra time-like dimensions
are non-dynamical and certainly metaphysical. Another example is an extra
space with no Killing vectors (see Aref'eva, Volovich in \cite{25}) which
prevents the appearance of ghost states in four dimensions.

Here we would like to suggest different possibility to eliminate tachyons
and ghosts from an effective four-dimensional theory. Suppose that bulk
space-time contains both time-like and space-like dimensions, i.e. the
metric of extra space is Lorentzian. Now we can make so called null space
reduction to lower dimensions demanding an extra subspace $M^{(p,q)}$ with $%
p $ space-like and $q$ time-like dimensions is actually a null space%
\footnote{%
Note that null space dimensional reduction is that what happens in $F$%
-theory (see Vafa in \cite{27}) where the BRST invariance demand the extra $%
M^{(1,1)}$ subspace of a $10+2=12$ dimensional type IIB string theory with
an $U(1)$ super-Maxwell field on the worldsheet to be a null space, so that $%
F$-theory in 12 dimensions with signature $(10,2)$ becomes dual to type IIB
string theory in 10 dimensions with signature $(9,1)$.}. If so, then the
physical states carry no non-zero momentum along $M^{(p,q)}$ subspace and so
all KK excitations will be confined on the light cone. Thus the physical
states (perhaps except of some additional zero mass modes) are the same as
in four-dimensional theory and the effective four-dimensional theory will be
tachyon-free. Another important outcome from the null reduction is that
since the physical states are essentially the same as in four dimensional
theory the physical laws will be also four-dimensional for all energy
scales. Say, $1/r^{2}$ 4-dimensional Newton's law will remain the same in
the null reduced higher-dimensional theory at any distances.

Below we explicitly demonstrate these features by considering a
6-dimensional space-time with one extra time-like and one extra space-like
dimensions, i.e. the space-time with signature (4,2). The null reduction is
realized by imposing additional discrete symmetry that prevents propagation
of the massive KK modes of graviton including those of tachyonic. In this
way we will avoide phenomenological difficulties comming from the matter
instability usually induced in theories with extra time-like dimensions.
Remarkable enough, that the same symmetry forced the cosmological constant
to be zero\footnote{%
For some earlier attempts to solve cosmological constant problem by
introducing extra time-like dimensions see Aref'eva, Dragovich, Volovich in 
\cite{25}; Linde in \cite{25}; \cite{21}.}.

\section{Intersecting branes: vacuum solutions.}

Let us consider 6-dimensional space-time $M^{(4,2)}$ with one extra
time-like dimension $\tau $ and one extra dimension $y$, i.e. space-time
with a signature $(4,2)$. Suppose that there are two branes with a
world-volume signature $(4,1)$ (``time brane'') and , $(3,2)$ ( ``space
brane'') embedded in $M^{(4,2)}$ with tensions $T_{\tau }$ and $T_{y}$,
respectively. The intersection of these branes, which we take to be at $%
(y=0,\tau =0)$ point for definiteness, is a 4-dimensional subspace (3-brane)
of $M^{(4,2)}$ with signature $(3,1)$ which can be identified with a visible
world. The relevant action describing such a set-up is: 
\begin{eqnarray}
S &=&\int d^{6}x\sqrt{\det g}\left( \frac{1}{\kappa _{6}^{2}}R-\Lambda
_{b}\right) -\int d^{6}x\sqrt{\det g}\,\left[ T_{\tau }\delta (\tau
)+T_{y}\delta (y)\right]  \nonumber \\
&=&\int_{M^{(4,2)}}d^{4}xdyd\tau \sqrt{\det g}\left( \frac{1}{\kappa _{6}^{2}%
}R-\Lambda _{b}\right)  \nonumber \\
&&-\int_{M^{(4,1)}}d^{4}xdy\sqrt{-\det g^{\tau =0}}T_{\tau
}-\int_{M^{(3,2)}}d^{4}xd\tau \sqrt{\det g^{y=0}}T_{y}.  \label{4}
\end{eqnarray}
Here $\kappa _{6}^{2}=16\pi M_{6}^{-4}$, where $M_{6}$ is the
six-dimensional fundamental scale of the theory and $\Lambda _{b}$ is a bulk
cosmological constant. The induced metrics on the branes, $g_{ab}^{\tau
=0}(a,b=\mu ,y)$ and $g_{\alpha \beta }^{y=0}(\alpha ,\beta =\mu ,\tau )$,
are defined as: 
\begin{eqnarray}
g_{ab}^{\tau =0} &=&g_{ab}(x^{\mu },y,\tau =0),  \nonumber \\
g_{\alpha \beta }^{y=0} &=&g_{\alpha \beta }(x^{\mu },y=0,\tau ),  \label{5}
\end{eqnarray}
where $g_{MN}$, $M,N=\mu (0,1,2,3),y,\tau $, is a six-dimensional metric. We
use metric with mostly positive signature $(-++++-)$. The field equations
followed from the above action (\ref{4}) are: 
\begin{equation}
R_{N}^{M}-\frac{1}{2}\delta _{N}^{M}R=\frac{\kappa _{6}^{2}}{2}T_{N}^{M},
\label{6}
\end{equation}
where the energy momentum tensor $T_{N}^{M}$ is expressed through the bulk
cosmological constant $\Lambda _{b}$ and brane tensions $T_{\tau }$ and $%
T_{y}$ as: 
\begin{equation}
T_{N}^{M}=-\Lambda _{b}\delta _{N}^{M}-\sqrt{\frac{-\det g^{\tau =0}}{\det g}%
}T_{\tau }\delta (\tau )\delta _{a}^{M}\delta _{N}^{a}-\sqrt{\frac{\det
g^{y=0}}{\det g}}T_{y}\delta (y)\delta _{\alpha }^{M}\delta _{N}^{\alpha }.
\label{7}
\end{equation}

We are looking for a static solution of the above equations (\ref{6}) that
respects 4-dimensional Poincare invariance in the $x^{\mu }$ direction. A
6-dimensional line element satisfying this anzatz can be written as: 
\begin{equation}
ds^{2}=A^{2}(y,\tau )\eta _{\mu \nu }dx^{\mu }dx^{\nu }+B^{2}(y,\tau
)dy^{2}-C^{2}(y,\tau )d\tau ^{2},  \label{8}
\end{equation}
where $\eta _{\mu \nu }$ is a 4-dimensional flat Minkowski metric. It is
more convenient, however, to perform the actual calculations within a
conformaly flat metric anzatz 
\begin{equation}
ds^{2}=A^{2}(z,\theta )\eta _{MN}dx^{M}dx^{N},  \label{9}
\end{equation}
which can be obtained from (\ref{8}) by the following coordinate
transformations\footnote{%
Generally this transformations do exist for the rather special cases.
However, they are actually valid in the case of the background solutions we
are interested in (see below).}: 
\begin{eqnarray}
dz &=&\frac{B}{A}dy,  \nonumber \\
d\theta &=&\frac{C}{A}d\tau .  \label{10}
\end{eqnarray}
Now using the well-known conformal transformation formulae for the Einstein
tensor $G_{N}^{M}=R_{N}^{M}-\frac{1}{2}\delta _{N}^{M}R$%
\begin{eqnarray}
\widetilde{G}_{MN} &=&G_{MN}+4\left( \nabla _{M}\ln A\nabla _{N}\ln A-\nabla
_{M}\nabla _{N}\ln A\right)  \nonumber \\
&&+4\eta _{MN}\left( \nabla ^{2}\ln A+\frac{3}{2}\left( \nabla \ln A\right)
^{2}\right) ,  \label{11}
\end{eqnarray}
we easily obtain: 
\begin{eqnarray}
G_{\nu }^{\mu } &=&\frac{2}{A^{2}}\left[ \left( \frac{A^{\prime }}{A}\right)
^{2}-\left( \frac{\dot{A}}{A}\right) ^{2}+2\left( \frac{A^{\prime \prime }}{A%
}-\frac{\ddot{A}}{A}\right) \right] \delta _{\nu }^{\mu },  \label{12} \\
G_{z}^{z} &=&\frac{2}{A^{2}}\left[ 5\left( \frac{A^{\prime }}{A}\right)
^{2}-\left( \frac{\dot{A}}{A}\right) ^{2}-2\frac{\ddot{A}}{A}\right] ,
\label{13} \\
G_{\theta }^{\theta } &=&\frac{2}{A^{2}}\left[ -5\left( \frac{\dot{A}}{A}%
\right) ^{2}+\left( \frac{A^{\prime }}{A}\right) ^{2}+2\frac{A^{\prime
\prime }}{A}\right] ,  \label{14} \\
G_{\theta }^{z} &=&-G_{z}^{\theta }=\frac{4}{A^{2}}\left[ 2\frac{\dot{A}%
A^{\prime }}{A^{2}}-\frac{\dot{A}^{\prime }}{A}\right] ,  \label{15}
\end{eqnarray}
where primes and overdots denote the derivatives with respect to space-like $%
z$ and time-like $\theta $ coordinates, respectively. Taking the conformal
factor (warp factor) in (\ref{9}) as\footnote{%
Turning back to the original coordinates $y$ and $\tau $ we will have: $%
A=\left( e^{k_{y}\left| y\right| }+e^{k_{\tau }\left| \tau \right|
}-1\right) ^{-1}$, $B=e^{k_{y}\left| y\right| }A$ and $C=e^{k_{\tau }\left|
\tau \right| }A.$} 
\begin{equation}
A=\frac{1}{k_{y}\left| z\right| +k_{\tau }\left| \theta \right| +1},
\label{16}
\end{equation}
one can easily check that non-diagonal elements (\ref{15}) of the Einstein
tensor vanish, $G_{\theta }^{z}=-G_{z}^{\theta }=0$, and thus $\left(
z\theta \right) $ Einstein's equations are satisfied identically, while the
remaining equations will be satisfied if the following relations are
fulfilled\footnote{%
While this paper was in preparation there appeared the paper \cite{23} in
the hep-archives where the same solution was considered. An earlier
presentation of the present work was given in \cite{24}.}: 
\begin{eqnarray}
k_{y}^{2}-k_{\tau }^{2} &=&-\frac{\kappa _{6}^{2}\Lambda _{b}}{10},
\label{17} \\
k_{y} &=&\frac{\kappa _{6}^{2}T_{y}}{4},  \label{18} \\
k_{\tau } &=&-\frac{\kappa _{6}^{2}T_{\tau }}{4}.  \label{19}
\end{eqnarray}
Here we will assume that the space brane has a possitive tension $T_{y}>0$,
while the time brane the negative one, $T_{\tau }<0$, so that both $k_{y}$
and $k_{\tau }$ are positive. In the opposite case, the negative (positive)
tension space (time) brane is expected to be unstable under the fluctuations
since in this case brane fluctuation modes show up as a ghost states in the
effective theory on the brane world-volume.

Taking away the ''time brane'' ($T_{\tau }=0$) one gets the 6-dimensional
version (with an extra time on the brane world-volume) of Ref. \cite{3} ,
where the bulk cosmological constant is negative, $\Lambda _{b}<0$, while in
the case $T_{y}=0$ one leads to the 6-dimensional version of the model of
Ref. \cite{20} with $\Lambda _{b}>0$. Thus, generally, depending on the
brane tensions $T_{y}$ and $T_{\tau }$, bulk space-time can be anti-de
Sitter $\left( \left| T_{\tau }\right| <\left| T_{y}\right| ,\Lambda
_{b}<0\right) $, de Sitter $\left( \left| T_{\tau }\right| >\left|
T_{y}\right| ,\Lambda _{b}>0\right) $ or Minkowskian $\left( T_{\tau
}=-T_{y},\Lambda _{b}=0\right) $. Remarkably, that in the latter case one
can observe an apparent discrete symmetry of the background solution (\ref
{16}-\ref{19}). Indeed, when one exchanges the extra time-like and
space-like dimensions, $\theta \longleftrightarrow z$ the background
solution (\ref{16}-\ref{19}) with $\Lambda _{b}=0$ remains untouched, while
the solution with $\Lambda _{b}<0$ goes to the one with $\Lambda _{b}>0$ and
vise versa. Thus if we demand that the Einstein equations (\ref{6}) are
invariant under the $\theta \longleftrightarrow z$ exchange than among the
solutions (\ref{16}-\ref{19}) the one with $\Lambda _{b}=0$ survives. The
fine tuning problem now is resolved since the above invariance demands $%
T_{\tau }=-T_{y}$ and ensures automatic cancellation of the 4-dimensional
cosmological constant.

One can worry that the above result just simply follows from the fact that
we have considered the anzatz (\ref{9}) (or equivalently (\ref{8})) where
the flatness of the 4-dimensional space-time of the intersection of branes
was already assumed. However, this is not the case. Indeed, one can start
from the more general anzatz by taking 
\begin{equation}
\widetilde{g}_{\mu \nu }=\left( 1-\frac{1}{4}H^{2}\eta _{\mu \nu }x^{\mu
}x^{\nu }\right) ^{-2}\eta _{\mu \nu }  \label{add1}
\end{equation}
instead of the flat 4-dimensional metric $\eta _{\mu \nu }$ in (\ref{9}).
Here $H$ is a ''Hubble constant'' on the intersection. Now the anzatz (\ref
{9}) with (\ref{add1}) instead of $\eta _{\mu \nu }$ describes maximally
symmetric 4-dimensional space-times of the intersection of branes, i.e. de
Sitter ($H^{2}>0$) or anti de Sitter ($H^{2}<0$) (the flat Minkowski case
considered above corresponds to $H=0$). Then the components of the Einstein
tensor (\ref{12}) and (\ref{13}), (\ref{14}) will be changed by the
additional term $+\frac{3H^{2}}{A^{2}}\delta _{\nu }^{\mu }$ and $+\frac{%
6H^{2}}{A^{2}}$, respectively, while (\ref{15}) will remain unchanged. It is
easy to see that the corresponding Einstein equations will remain invariant
under the discrete symmetry $\theta \longleftrightarrow z$ if and only if $%
H=0$ and $\Lambda _{b}=0$. This can be easily understood from the fact that
the origin for the non-zero Hubble constant is a non-zero 4-dimensional
cosmological constant on the intersection of branes which in turn is indeed
forbidden if one demands that the theory is invariant under the discrete
symmetry imposed above.

Of course, in general, for an arbitrary metric $g_{MN} $ the invariance we
have imposed does not take place. In what follows we require here that such
an invariance holds for all perturbations around the background solution as
well. In other words, we restrict to consider the manifolds which are
isometric under the discrete symmetry transformations $\theta
\longleftrightarrow z$. So, the above invariance can be viewed as a
constraint imposed on the system described by the action (\ref{4}) which
holds for the special class of metrics $g_{MN}$ including the background one
given by (\ref{9},\ref{16}-\ref{19}) with $\Lambda _{b}=0$\footnote{%
On the language of the initial action (\ref{4}) the vanishing of
cosmological constant can be related to the following symmetry reasons. One
can consider a discrete transformation changing the signature of the metric, 
$g_{MN}\to -g_{MN}$ (and hence $R\to -R$). The first term in the action (\ref
{4}) changes the sign if $\Lambda _{b}=0$. (In fact, in theories with even 
number of space-time dimensions the bulk comsological
constant can be forbidden by assuming that
when the metric changes the signature also the action
changes the sign. E.g. in the usual $1+3$ case
the change $g_{\mu\nu} \to -g_{\mu\nu}$ is
equivalent to the changing the signature $(+,-,-,-)$
by the signature $(-,+,+,+)$).
If along this transformation also the $\tau$ and $y$
branes are exchanged, then the whole action changes
sign, $S \to -S$, if $T_\tau = - T_y$.}. Notice
that the vanishing of the bulk cosmological constant, $\Lambda_b=0$, and the
relation $T_{\tau }=-T_{y}$ emerge merely from the discrete symmetry imposed
and are not consequence of any fine-tuning. As we will see this symmetry
leads at the same time to the resolution of the problem of tachyonic states.

\section{Linearized perturbations.}

To examine the gravity induced by a matter source localized on the
4-dimensional intersection of branes let us consider the linearized
perturbations around the background metric (\ref{9},\ref{16}-\ref{19}).
Taking 
\begin{equation}
g_{MN}=A^{2}(z,\theta )\left( \eta _{MN}+\gamma _{MN}\right)  \label{20}
\end{equation}
and keeping linear in $\gamma _{MN}$ terms only in (\ref{6}) we get: 
\[
-\frac{1}{2}\partial _{K}\partial ^{K}\gamma _{MN}-\frac{1}{2}\partial
_{M}\partial _{N}\gamma _{K}^{K}+\partial ^{K}\partial _{(M}\gamma _{N)K}+%
\frac{1}{2}\eta _{MN}\left( \partial _{K}\partial ^{K}\gamma
_{L}^{L}-\partial ^{K}\partial ^{L}\gamma _{KL}\right) + 
\]
\begin{equation}
2\left[ \left( 2\partial _{(M}\gamma _{N)K}-\partial _{K}\gamma _{MN}-\eta
_{MN}\left( 2\partial ^{L}\gamma _{KL}-\partial _{K}\gamma _{L}^{L}+2\gamma
_{KL}\partial ^{L}+3\gamma _{KL}n^{L}\right) \right) \right] n^{K}=0
\label{21}
\end{equation}
where $n_{K}\equiv \partial _{K}\ln A=\left( 0,0,0,0,\frac{A^{\prime }}{A},%
\frac{\dot{A}}{A}\right) $ and indices are raised and lowered by the $6$%
-dimensional flat metric $\eta _{MN}$. We work with so-called
Randall-Sundrum (RS) gauge \cite{3,20,28} which is defined as a
4-dimensional transverse traceless gauge 
\begin{equation}
\partial ^{\mu }\gamma _{\mu \nu }=0,\hspace{0.5cm}\gamma _{\mu }^{\mu
}\equiv \gamma =0  \label{22}
\end{equation}
along with an extra conditions 
\begin{equation}
\gamma _{M\tau }=\gamma _{My}=0.  \label{23}
\end{equation}
Although RS gauge (\ref{22},\ref{23}) becomes inconsistent with equations of
motion when one considers an extra matter sources beyond those given by
branes itself \cite{29}, for the discussion of the gravity localization on
the intersection of branes it is rather convenient and we will keep it here
and return to this point later\footnote{%
For some alternative gauges as well as discussions on subtleties in the RS
gauge choice see \cite{29,30}}.

In the above gauge (\ref{22},\ref{23}) one left with 2 physical massless
degrees of freedom on the intersection, which corresponds to just 2
polarization states of the 4-dimensional graviton and the equations (\ref{21}%
) simplify significantly to become: 
\begin{equation}
\partial _{M}\partial ^{M}\gamma _{\mu \nu }+4\partial _{M}\gamma _{\mu \nu
}n^{M}=0.  \label{24}
\end{equation}
Boundary conditions on $\gamma _{\mu \nu }$ can be deduced by integration of
(\ref{24}) from just below to just above of the time and space branes
resulting in the Darmois-Israel matching conditions \cite{31}\footnote{%
Actually from the Darmois-Israel formalism one can get in general another
type of boundary conditions, which tell us that the derivatives of the
metric are just continuous $\gamma _{\mu \nu }^{\prime }\mid
_{z=0^{+}}=\gamma _{\mu \nu }^{\prime }\mid _{z=0^{-}},$ $\dot{\gamma}_{\mu
\nu }\mid _{\theta =0^{+}}=\dot{\gamma}_{\mu \nu }\mid _{\theta =0^{-}}$,
but not necessarily zero as in (\ref{25}). Here, following to \cite{3}, we
also assume that similar to the background metric the perturbations are also
even under the discrete transformations $z\rightarrow -z$ and $\theta
\rightarrow -\theta .$ Then the only boundary conditions consistent with
these symmetries are those given in (\ref{25}).}: 
\begin{equation}
\gamma _{\mu \nu }^{\prime }\mid _{z=0}=0,\hspace{0.5cm}\dot{\gamma}_{\mu
\nu }\mid _{\theta =0}=0  \label{25}
\end{equation}
To proceed further, we separate the intersection world-volume and extra
space-time coordinates in (\ref{24}) setting $\gamma _{\mu \nu }=h_{\mu \nu
}(x^{\sigma })\Psi _{m}(z,\theta )$. Then the Eqs.(\ref{24}) are splitted
as: 
\begin{equation}
\partial _{\sigma }\partial ^{\sigma }h_{\mu \nu }=m^{2}h_{\mu \nu },
\label{26}
\end{equation}
\begin{equation}
\left( \partial _{y}\partial ^{y}-\partial _{\tau }\partial ^{\tau
}+4n_{y}\partial _{y}-4n_{\tau }\partial _{\tau }\right) \Psi _{m}(z,\theta
)=-m^{2}\Psi _{m}(z,\theta ),  \label{27}
\end{equation}
where $m^{2}$ is a separation constant. The eqs. (\ref{26},\ref{27})
describe the propagation of a free massive spin-2 particle in the
intersection world-volume with a mass $m^{2}$ and the wave function $\Psi
_{m}(z,\theta )$. Although we are not able to solve the equation (\ref{27})
explicitly in the general case, but it is obvious that one has normalizable
solutions for the wave function $\Psi _{m}(z,\theta )$ corresponding to the
mass eigenvalues in the whole range $-\infty <m^{2}<+\infty $. This readily
follow from the fact that the normalize massive KK excitations along the
time-like extra dimension are actually tachyonic ($m^{2}<0$) \cite{20} while
those of along the space-like extra dimension are bradionic ($m^{2}>0$) \cite
{3}. Finally, in general case, one expects also a collection of infinitely
many massless gravitons propagating in the 4-dimensional world-volume. One
of them is a ''true'' zero mode localized on the intersection, while others
are the excitations along the light-cone in extra space-time. Obviously,
with such a spectrum of KK-states one can not have any satisfactory
phenomenology on the intersection world-volume. Besides the already
mentioned problem with tachyonic KK-states of graviton \cite{20}, one could
worry about the infinitely degenerate massless gravitons as well, since they
also might bring inconsistencies when one goes beyond the linearized
approximation \cite{32}.

Now let us proceed to the nul space dimensional reduction by requiring that
eqs. (\ref{26},\ref{27}) are invariant under the interchange of extra
time-like and extra space-like coordinates, $\theta \longleftrightarrow z$,
as it was proposed in the previous section. This symmetry demand vanishing
of $m^{2}$ in (\ref{26},\ref{27}), $m^{2}=0$, and the eqs. (\ref{26},\ref{27}%
) now become: 
\begin{equation}
\partial _{\sigma }\partial ^{\sigma }h_{\mu \nu }=0  \label{28}
\end{equation}
\begin{equation}
\left( \partial _{y}\partial ^{y}-\partial _{\tau }\partial ^{\tau
}+2kA\left( \delta (\theta )-\delta (z)\right) \right) \psi (z,\theta )=0.
\label{29}
\end{equation}
Here we set $\psi (z,\theta )=A^{2}\Psi _{0}(z,\theta )$ in order to
canonically normalize the kinetic term in (\ref{29}) and $k_{y}=k_{\tau
}\equiv k$. Now, eqs. (\ref{28},\ref{29}) describe the propagation of
massless spin-2 particle localized on the 4-dimensional intersection region.
Indeed, the only normalizable solution which is consistent with boundary
conditions (\ref{25}) is: 
\begin{equation}
\psi (z,\theta )=\sqrt{\frac{3}{2}}\frac{k}{\left( k\left( \left| z\right|
+\left| \theta \right| \right) +1\right) ^{2}}  \label{30}
\end{equation}
where we have properly normalized the wave function, $\int \int dzd\theta
\left| \psi (z,\theta )\right| ^{2}=1$. Thus the discrete symmetry $\theta
\longleftrightarrow z$ not only provides automatic cancelation of the
effective cosmological constant on the 4-dimensional intersection, but also
singles out the zero mode solution (\ref{30}), for the graviton. This in
turn means that gravity is fully localized on the 4-dimensional intersection
and the Newton law at all distances there is just ordinary 4-dimensional one. 
Indeed the Newton potential of the two point-like mass $M_{1}$ and $M_{2}$
placed on the intersection at a distance $r$ from each other is: 
\begin{equation}
V(r)=G_{N}^{(6)}\frac{M_{1}M_{2}}{r}\left| \psi (0,0)\right| ^{2}=G_{N}^{(4)}%
\frac{M_{1}M_{2}}{r}  \label{31}
\end{equation}
where the effective 4-dimensional Newton constant $G_{N}^{(6)}\equiv \frac{%
\kappa _{4}^{2}}{2}=8\pi M_{4}^{-2}$ is related to the 6-dimensional one $%
G_{N}^{(6)}\equiv \frac{\kappa _{6}^{2}}{2}=8\pi M_{6}^{-4}$ through the
following relation: 
\begin{equation}
G_{N}^{(4)}=\frac{3k^{2}}{2}G_{N}^{(6)}  \label{32}
\end{equation}
Note also that, while the effects of extra dimensions are essentially hidden
for the 4-dimensional observer, they show up in the non-trivial warp factor $%
A$ (\ref{16}). Thus the solution to the Plank/TeV scale hierarchy problem is
possible in our scheme in the spirit proposed in \cite{6}, i.e. by placing
the TeV 3-brane with Standard Model particles localized on it in the bulk at
an appropriate distance from the intersection. Finally, the extension of the
above scheme to more extra dimensions is also possible, although it requires
more sophisticated extra symmetries to remove tachyonic KK modes from the
effective 4-dimensional theory and to ensure natural (without fine tuning)
vanishing of the cosmological constant.

\section{Gravity on the intersection in the Minkowski bulk.}

Here we would like to discuss the null space dimensional reduction in the
case of the flat Minkowski bulk space-time. From the preceding sections, at
first glance, it seems that KK modes are confined on the light cone even in
the case of a Minkowski bulk space-time and thus 4-dimensional observer will
still obtain $\frac{1}{r^{2}}$ Newton's law on the intersection. Then,
leaving aside the solution to the hierarchy problem via warped
compactification, one can ask: do we really need to warp up or compactify
the extra dimensions in order to have consistent 4-dimensional physics on
the intersection region? The answer seems would be positive for all
interactions except of gravity. Indeed, to have a consistent 4-dimensional
gravity on the intersection it is not sufficient to correctly reproduce the
Newton law, but one should ensure that the gravity on the intersection is
just a tensor-type, i.e. an extra polarization states are actually decoupled
from the 4-dimensional massless graviton. If it is not a case then one fails
to explain some familiar gravitational experiments such as the experiments
on the bending of light in the gravitational field of the Sun and the
experiments measured precession rate of the Mercury orbit.

These extra polarization states show up in the tensor structure of the
propagator of massless graviton. Basically this is related to the well-known
fact that the massless limit of the massive graviton is actually
discontinuous in the flat background space-time \cite{33}. So one should
evaluate full propagator including the tensor structure as well. Here we
come to the point the discussion on which we have postponed in the previous
section. As we have mentioned there, the RS gauge (\ref{20},\ref{23})
becomes inconsistent when one consider an extra matter sources say localized
on the intersection of the branes. One can relax the traceless condition in (%
\ref{20}), but now extra scalar polarization state appear in the spectrum of
the massless states on the 4-dimensional intersection. Thus one can worry
that this scalar polarization state can not be removed from the physical
spectrum and the gravity on the 4-dimensional intersection is scalar-tensor
type. As it is shown in \cite{29}, actually this is not the case in the
original RS model \cite{3}. The extra scalar can indeed be gauged away. The
physical reason for this is that in the RS model one has a zero mode
graviton localized on the brane for which one expect to have the usual
4-dimensional massless propagator. Since in the case considered in the
previous section the graviton zero mode is also localized on the
intersection, the same procedure used in \cite{29} can be applied to show
that the gravity on the intersection is just the Einstein-type one.

Now turning to the case of the Minkowski bulk, it is obvious that the extra
polarization states can not be gauged away, since gravity is allowed to
freely propagate in the infinite Minkowski bulk space-time. Although we can
hide all massive KK-states through the null space dimensional reduction, so
that the scalar part of the propagator of graviton will be just as for the
massless 4-dimensional one, but the tensor structure of the propagator is
expected to be higher-dimensional, due to the contributions of the massless
scalar modes. Note once again that this point is peculiar to the
gravitational interactions. Say gauge fields, living in the null reduced
Minkowski bulk, can correctly reproduce 4-dimensional physics on the
intersection.

\section{Conclusions.}

We discussed the cosmological constant problem within the Brane World
scenario with extra time-like dimension. Particularly, we considered the
intersecting brane configuration in the 6-dimensional space-time with one
extra time-like and one extra space-like dimensions. Among the possible
warped background solutions to the Einstein equations we have found one with
vanishing cosmological constant which is invariant under the discrete
interchange of extra time-like and extra space-like dimensions. This simple 
symmetry can be suitably generalized (see footnote 9) to ensure vanishing 
of the bulk cosmological constant, $\Lambda _b=0$, and the realation 
$T_{\tau }=-T_{y}$ in the original action (\ref{4}) and thus to single out the 
desired background solution with flat 4-dimensional intersection of branes. 
On the other hand, it is remarkable that the same symmetry leads to the fully localized
gravity on the 4-dimensional intersection world-volume. Thus the problem of
natural (without fine tuning) vanishing of the cosmological constant and the
stability problem related with extra time-like dimension are solved
simultaneously. 

Clearly, several questions have to be answered before the above proposal can be 
considered as a candidate solution to the cosmological constant problem. Among them 
are the localization of Standard Model fields on the 4-dimensional  
intersection and the origin of the discrete symmetry imposed, which in the given set-up 
perhaps looks rather artificial. One can hope to find an answers to these 
questions at a more fundamental level. In this respect it is interesting to 
investigate whether the
above or similar constructions can be obtained as a low energy limit of more
fundamental string theory.

\paragraph*{Acknowledgments.}

We thank Gia Dvali, Gregory Gabadaze, Merab Gogberashvili, Zurab Kakushadze
and Dimitri Polyakov for useful discussions. A.B.K. would like also to
gratefully acknowledge stimulating research atmosphere at Gran Sasso Summer
Institute ''Dark matter and Supersymmetry'' where the part of this work was
done. The research of M.C., A.B.K and Z.H.Y. was supported by the Academy of
Finland under the Research Project No. 163394 and that of Z.B. was partially
supported by the MURST research grant ''Astroparticle Physics''.\newpage


\end{document}